\documentclass[]{interact}

\usepackage{epstopdf}
\usepackage[caption=false]{subfig}
\usepackage[numbers,sort&compress]{natbib}
\bibpunct[, ]{[}{]}{,}{n}{,}{,}

\theoremstyle{plain}

\theoremstyle{definition}

\theoremstyle{remark}

\begin{document}


\title{The interplay between a pseudogap and superconductivity in a two-dimensional Hubbard model}

\author{
\name{L. F. Sampaio\textsuperscript{a}, E. J. Calegari\textsuperscript{a}\thanks{CONTACT E. J. Calegari. Email: eleonir@ufsm.br}, J.J. Rodr\'{\i}guez-N\'u\~nez\textsuperscript{b}, A. Bandyopadhyay\textsuperscript{a} and R. L. S. Farias\textsuperscript{a}}
\affil{\textsuperscript{a}Departamento de F\'{\i}sica, Universidade Federal de Santa Maria, 97105-900, Santa Maria, RS, Brazil; \textsuperscript{b} Laboratorio de Superconductividad Computacional ($SUPERCOMP$), De\-par\-ta\-men\-to de F\'isica - FACYT,
Universidad de Carabobo, Valencia, Venezuela
}}

\maketitle

\begin{abstract}

Strongly correlated electrons systems may exhibit a variety of interesting phenomena, for instance, superconductivity and pseudogap, as is the case of cuprates and pnictides. In strongly correlated electron systems, it is
considered essential to understand, not only the nature of the pseudogap, but also the relationship between superconductivity and the pseudogap. In order to address this question, in the present work, we investigated a one-band Hubbard model treated by the Green’s function method within an n-pole approximation.
In the strongly correlated regime,  antiferromagnetic correlations give rise to
nearly flat band regions in the nodal points of the  quasiparticle bands. As a consequence, a pseudogap emerges at the antinodal points of the Fermi surface. 
The obtained results indicate that the same antiferromagnetic correlations responsible for a pseudogap, can also favor superconductivity, providing an increase in the superconducting critical temperature T$_c$.

\end{abstract}

\begin{keywords}
flat bands;   superconductivity; antiferromagnetic correlations; critical temperature; pseudogap
\end{keywords}

\section{Introduction}

Since the discovery of superconductivity in cuprates  \cite{Bednorz} more than three decades ago, superconductivity in strongly correlated electron systems
is one of the most important subjects
in condensed matter physics. Although a lot of effort
has been devoted to understanding the mechanisms of superconductivity in these systems, it is still  an open problem.
In correlated electron systems, superconductivity is closely related to the underlying band structures \cite{aoki,ino}. For instance, systems which present a flat band, or even a partially flat band, may exhibit high temperature transitions \cite{aoki,edwin,edwin1,Melnikov}. Indeed, systems with flat bands at the Fermi energy $E_F$ give rise to a large density of states (DOS) at $E_F$,
which implies that many electrons contribute to
low energy phenomena, for example,
superconductivity \cite{Kauppila,Markiewicz}.

Another important feature related to some strongly correlated electron systems, is the  pseudogap phenomenon \cite{Cyr,proust,solov,gull,sakai}, which is a partial gap in the spectral function that opens at the Fermi energy.
There are different proposals to explain the origin of the pseudogap on correlated superconductors \cite{proust,solov}. For instance,  we can mention  charge density waves (CDW) \cite{Borisenko}, pre-formed pairs \cite{yanase} and superconducting phase fluctuations \cite{Emery,Loktev}, as possible sources of the pseudogap.
On the other hand, there are experimental \cite{Cyr,Naqib,Chan} and theoretical \cite{Harrison,Morinari,valerii,Korshunov,Kuzmin,avella,Thomas,kai,Wei,Macridin,Kancharla,Gunnarsson,Senechal,Vilk} results
suggesting that short-range antiferromagnetic correlations present in the low doping regime, may be the source of a pseudogap.
Although the pseudogap occurs in the normal state, researchers believe it
essential to identify
its origin
in order to unravel the mechanisms behind the superconductivity in correlated superconductors, such as cuprates and pnictides \cite{moon,jesus2022,jesus2023}. Therefore, in these systems it is really important not only to unveil the nature of the pseudogap, but also to understand the interplay between superconductivity and pseudogap.

In order to address this important question, in the present work,  we consider a two-dimensional Hubbard model \cite{ref15} in order to investigate the relationship between a pseudogap and superconductivity, in a scenario in which short-range antiferromagnetic correlations are the source of a pseudogap.
From the theoretical viewpoint, the one-band Hubbard model
has been widely investigated as a
model that
contains the basic ingredients
necessary to describe the physics of some strongly correlated superconductors \cite{ edwin,avella,beenen,Fedor}.
In this work, the Hubbard model has been treated by using the Green's function technique combined with an n-pole approximation \cite{beenen,Roth,calegari2005,calegari2011,diovana}, which allows to obtain a close set of decoupled Green's functions. The  quasiparticle bands
of the Green's functions, have a band shift $W_{\vec{k}\sigma}$ which
includes important correlation functions that
are neglected in the mean field and Hubbard I approximations \cite{ref15}.  One of the most important correlation functions present in the band shift, is the spin-spin correlation function  $\langle \vec{S}_i\cdot\vec{S}_j\rangle$, which is associated
with antiferromagnetic correlations \cite{beenen},
the
source of a pseudogap in the present scenario \cite{Harrison,avella,calegari2011,diovana}. In the original n-pole approximation applied to calculations on the Hubbard model by Roth \cite{Roth}, the band shift was calculated by considering $t_{ij}=t$ (where $t_{ij}$ is the hopping between nearest-neighbors) for the $z$ nearest neighbors. Nevertheless, such a simplification does not adequately capture the momentum dependence of the band shift,  mainly affecting the spin-spin correlation function responsible for the pseudogap.
In the present work, this original version \cite{Roth} of the n-pole approximation, will be called the Roth1 method. In order to treat the momentum dependence of the band shift in a more appropriate way, 
we follow the procedure considered in references \cite{calegari2011, diovana},
called here,
the Roth2 method.

In the Roth2 method, the effect of the antiferromagnetic correlations associated with the spin-spin correlation function present in the band shift, may produce a quasiparticle band with a nearly flat region at the antinodal points $(0,\pm \pi)$ and $(\pm\pi,0)$. Such a flat region is responsible not only  for the emergence of the pseudogap, but also  for the enhancement of the superconducting critical temperature, due to the large DOS at the Fermi energy \cite{Kauppila,Markiewicz}.

There are some shortcomings in the n-pole approximation. In order to evaluate the correlation functions present in the band shift, Roth introduced a set of auxiliary operators \cite{Roth}. However, there is not a single choice for this set of operators, which may lead to inappropriate choices. For the purpose of evaluating the choice of the auxiliary operators considered by Roth, in reference \cite{beenen} the quasiparticle bands obtained with Roth’s approximation and those from quantum Monte Carlo calculations \cite{bulut} were compared and a good agreement between the results was verified. Moreover, in a recent work \cite{Haurie}, the authors developed a detailed investigation comparing the two solutions (COM1 and COM2) from the composite operator method (COM) and Roth’s solution. It was observed that, although Roth’s solution can violate the Pauli principle, it exhibits Fermi surfaces typical of strongly correlated materials such as cuprates.

The superconductivity is taken into account in the BCS \cite{BCS,jesus350} sense, and  the repulsive Coulomb interaction is introduced through the Hubbard model following the methodology considered in references \cite{jesus350,Tifrea2003,moca,calegari2016}.
Although the repulsive one-band Hubbard model would be enough to investigate  superconductivity with $d_{x^2-y^2}$-wave pairing in strongly correlated electron systems, the repulsive interaction term makes it a complex task to treat the equations of motion of Green's functions in the n-pole approximation. On the other hand, the addition of the BCS-type term to the one-band Hubbard model significantly simplifies the problem because, as will be discussed in section \ref{formalism}, the attractive term is treated in the mean field level while the correlated Green’s function $G_N$ is  obtained only in the normal state. Even though we are aware that this procedure may cause the loss of some information,
this procedure allows us to carry out an investigation of how the antiferromagnetic correlations responsible for the pseudogap can affect quantities such as the superconducting order parameter and the superconducting critical temperature. In addition, we verified that the numerical results for the quasiparticle bands in the superconducting state (not shown here) agree well with those reported by Beenen and Edwards\cite{beenen} and recently  by Haurie et al \cite{Haurie}. The behavior of the superconducting order parameter as a function of electron density also agrees with the data reported in the above-mentioned references.

This paper is organized as follows. In section \ref{formalism}, the model is presented, and equations for the superconducting gap and the critical temperature, are introduced. In section \ref{numresults}, we present the numerical results, while section \ref{conclusions} is devoted to conclusions.

\section{The formalism}
\label{formalism}
The Hamiltonian of the model \cite{jesus350,kwon,Chen}, is given by:
\begin{equation}
\hat{\cal{H}}=\hat{H}_U+\hat{H}_{PAR}
\label{eqH}
\end{equation}
in which $\hat{H}_U$ is the  two-dimensional one-band Hubbard model \cite{ref15}:
\begin{equation}
\hat{H}_U=\sum_{i,j,\sigma}t_{ij}\hat{c}^{\dagger}_{i,\sigma}\hat{c}_{j,\sigma}+\frac{U}{2}\sum_{i,\sigma}
\hat{n}_{i,-\sigma}\hat{n}_{i,\sigma}  -\mu\sum_{i,\sigma}\hat{c}^{\dagger}_{i,\sigma}\hat{c}_{i,\sigma}.
\label{hu1}
\end{equation}
The  creation(destruction) operators $\hat{c}^{\dagger}_{i,\sigma}(\hat{c}_{i,\sigma})$ can create(destroy) an electron with spin $\sigma$ on the lattice  site $i$, while $\hat{n}_{i,\sigma}=\hat{c}^{\dagger}_{i,\sigma}\hat{c}_{i,\sigma} $ is the number operator.  The first term in $\hat{H}_U$ describes the hopping of the electrons through the lattice sites while
the second term considers the repulsive Coulomb interaction between two electrons with opposite spins located at the same site $i$. The third term considers the chemical potential $\mu$.  For a simple square lattice, the dispersion relation is
 \begin{equation}
 \varepsilon_{\vec{k}}=2t[\cos{(k_xa)}+\cos{(k_ya)}]-4t'\cos{(k_xa)}\cos{(k_ya)},
 \end{equation}
 where $a$ is the lattice parameter and $t$ and $t'$ are the hopping amplitudes for the first and second nearest-neighbors, respectively.

The paring term \cite{tinkham}
\begin{equation}
\hat{H}_{PAR}=\sum_{\vec{k},{\vec{k}}^{'}}V_{\vec{k},{\vec{k}}^{'}}\hat{c}^{\dagger}_{\vec{k}, {\uparrow}} \hat{c}^{\dagger}_{-\vec{k},{\downarrow}}\hat{c}_{-\vec{k}^{'},{\downarrow}} \hat{c}_{\vec{k}^{'},{\uparrow}}
\label{bcs}
\end{equation}
is treated in the BCS \cite{BCS,jesus350} level, and  we leave unspecified the origin of the attractive interaction $V_{\vec{k}, \vec{k}^{'}}$.

The present methodology, which allows to obtain the correlated Green's functions, consists in, first, calculating the uncorrelated Green's functions for the superconducting state.
Therefore, initially we consider  $U=0$ and calculate the equations of motion for the Hamiltonian of Eq. (\ref{eqH}), at a mean-field level, where the effects of temperature are included through the imag\-inary-time formalism \cite{lundberga}. In this case:
\begin{eqnarray}
G^{-1}_{0}(\vec{k},i\omega_{n})G(\vec{k},i\omega_n)-\Delta_{\vec{k}}F^{\dagger}(\vec{k},i\omega_n)=1 \nonumber\\
G^{-1}_{0}(\vec{k},i\omega_{n})F^{\dagger}(\vec{k},i\omega_{n})+\Delta^{*}_{\vec{k}}G(\vec{k},i\omega_{n})=0
\label{eq4}
\end{eqnarray}
where  $G_0(\vec{k},i\omega_n)=(i\omega_n-\varepsilon_{\vec{k}})^{-1}$ is the uncorrelated green's function for the normal state, while  $G$ and $F$ are the normal and the anomalous uncorrelated Green's functions, respectively.
The quantities $\Delta_{\vec{k}}$ and $i\omega_n=(2n+1)\frac{i\pi}{\beta}$ are the superconducting gap and the fermionic Matsubara frequencies, respectively. The inverse temperature is $\beta=\frac{1}{k_BT}$, with $T$ and $k_B$ denoting the temperature and Boltzmann constant, respectively. In order to taken into account correlations \cite{jesus350,Tifrea2003,moca,calegari2016}, the following substitution has been considered:
\begin{equation}
    G_{0}(\vec{k},i\omega_{n})\rightarrow  G_N(\vec{k},i\omega_{n})
\end{equation}
in which $G_N(\vec{k},i\omega_{n})$ is the correlated one-particle Green's function for the normal state. It has been assumed that the presence of correlations does not significantly affect the BCS formalism \cite{jesus350,jesusIJMPB}. Solving the set of Eqs. (\ref{eq4}), we obtain the correlated Green's functions for the superconducting state
\begin{equation}
G(\vec{k},i\omega_{n})=\frac{G^{-1}_{N}(\vec{k},i\omega_{n})}{|G_{N}(\vec{k},i\omega_{n})|^{-2}+|\Delta_{\vec{k}}|^{2}}
\end{equation}
and
\begin{equation}
F^{\dagger}(\vec{k},i\omega_{n})=\frac{-\Delta^{*}_{\vec{k}}}{|G_{N}(\vec{k},i\omega_{n})|^{-2}+|\Delta_{\vec{k}}|^{2}}.
\label{Fk}
\end{equation}

The normal correlated Green's function $G_N$ for the model given in Eq. (\ref{hu1}), has been calculated within an n-pole approximation (Roth method) \cite{beenen,Roth,calegari2005,calegari2011,diovana}
\begin{equation}
   G_N(\vec{k},i\omega_{n})=\frac{Z_{1\vec{k}}}{i\omega_n-E_{1\vec{k}}}+ \frac{Z_{2\vec{k}}}{i\omega_n-E_{2\vec{k}}}
   \label{GN}
\end{equation}
with the spectral weights given by:
\begin{equation}
    Z_{1\vec{k}}=\frac{X_{\vec{k}}+U(1-2n_{-\sigma})-\varepsilon_{\vec{k}}+W_{\vec{k},-\sigma}}{2X_{\vec{k}}} ~~~~\hbox{and}~~~~Z_{2\vec{k}}=1-Z_{1\vec{k}}.
\end{equation}
The quasiparticle bands are
\begin{equation}
    E_{1\vec{k}}=\frac{  U+\varepsilon_{\vec{k}}+W_{\vec{k},-\sigma}-X_{\vec{k}}}{2}-\mu ~~~~\hbox{and}~~~~E_{2\vec{k}}=E_{1\vec{k}}+X_{\vec{k}},
    \label{eqE1k}
\end{equation}
where
\begin{equation}
   X_{\vec{k}}=\sqrt{(U-\varepsilon_{\vec{k}}+W_{\vec{k},-\sigma})^2+4n_{-\sigma}U(\varepsilon_{\vec{k}}-W_{\vec{k},-\sigma})}.
\end{equation}

The Roth method maintains important correlation functions that are neglected in the mean-field \cite{jesus350} and Hubbard I \cite{ref15} approximations. In the Roth2 method \cite{calegari2011,diovana}, such correlation functions are present in the band shift $W_{\vec{k},-\sigma}$, which is given by,
\begin{align}
n_{-\sigma}(1-n_{-\sigma})W_{\vec{k},-\sigma}=S_{-\sigma}+\sum_{\vec{q}}\varepsilon_{\vec{k}-\vec{q}}
F_{\vec{q},\sigma}
\end{align}
with
\begin{equation}
S_{-\sigma}=-\sum_{j\neq i}t_{ij}\langle\hat{c}^{\dagger}_{i,\sigma}\hat{c}_{j,\sigma}(1-\hat{n}_{i,-\sigma}-\hat{n}_{j,-\sigma})\rangle.
\label{eqS}
\end{equation}
The quantity $F_{\vec{q},\sigma}$, is given in terms of the Fourier transform of the following correlation functions \cite{beenen,calegari2011}
\begin{eqnarray}
\label{eqS1}
S^{(1)}_{i,j,\sigma}=\frac{1}{4}(\langle \hat{N}_{j}\hat{N}_{i}\rangle-\langle \hat{N}_{j}\rangle\langle\hat{N}_{i}\rangle),\\
\label{eqS2}
S^{(2)}_{ij,\sigma}=\langle \vec{\hat{S}}_{j}\cdot\vec{\hat{S}}_{i}\rangle,\\
\label{eqS3}
S^{(3)}_{ij,\sigma}=-\langle\hat{c}_{j,\sigma}^{\dagger} \hat{c}_{j,-\sigma}^{\dagger}\hat{c}_{i,-\sigma}\hat{c}_{i,\sigma}\rangle
\end{eqnarray}
where $\hat{N}_j=\hat{n}_{j\sigma}+\hat{n}_{j-\sigma}$, is the total number operator and the dispersion relation is
\begin{equation}
\varepsilon_{\vec{k}-\vec{q}}=\frac{1}{L}\sum_{j\neq l}e^{i(\vec{k}-\vec{q})\cdot(\vec{R}_{j}-\vec{R}_{l})}t_{lj}.
\end{equation}
Here $L$ is the number of lattice sites in the system.

The correlation functions present in Eqs. (\ref{eqS}), (\ref{eqS1}), (\ref{eqS2}) and (\ref{eqS3}), have been calculated following the procedure proposed by Roth \cite{beenen, Roth,calegari2011}.   In particular, the spin-spin correlation function $\langle \vec{S}_j\cdot\vec{S}_i\rangle$ present in $S^{(2)}$ plays an important role  because it is related to  antiferromagnetic correlations  \cite{beenen}  that are one of the sources of a pseudogap in the density of states \cite{Harrison,Morinari,valerii}.

The gap equation is obtained from the anomalous Green function given in Eq. (\ref{Fk}), and is written as:
\begin{equation}
\Delta_{\vec{k}}=-\sum_{\vec{k^{'}}}V_{\vec{k},\vec{k^{'}}}\left(\frac{1}{\beta}\sum_{n}\frac{\Delta(\vec{k'})}{|G_{N}(\vec{k},i\omega_{n})|^{-2} +|\Delta_{\vec{k'}}|^{2}}\right).
\label{ffgg11}
\end{equation}
\begin{figure}[!ht]
\begin{center}
\leavevmode
\includegraphics[angle=0,width=9cm]{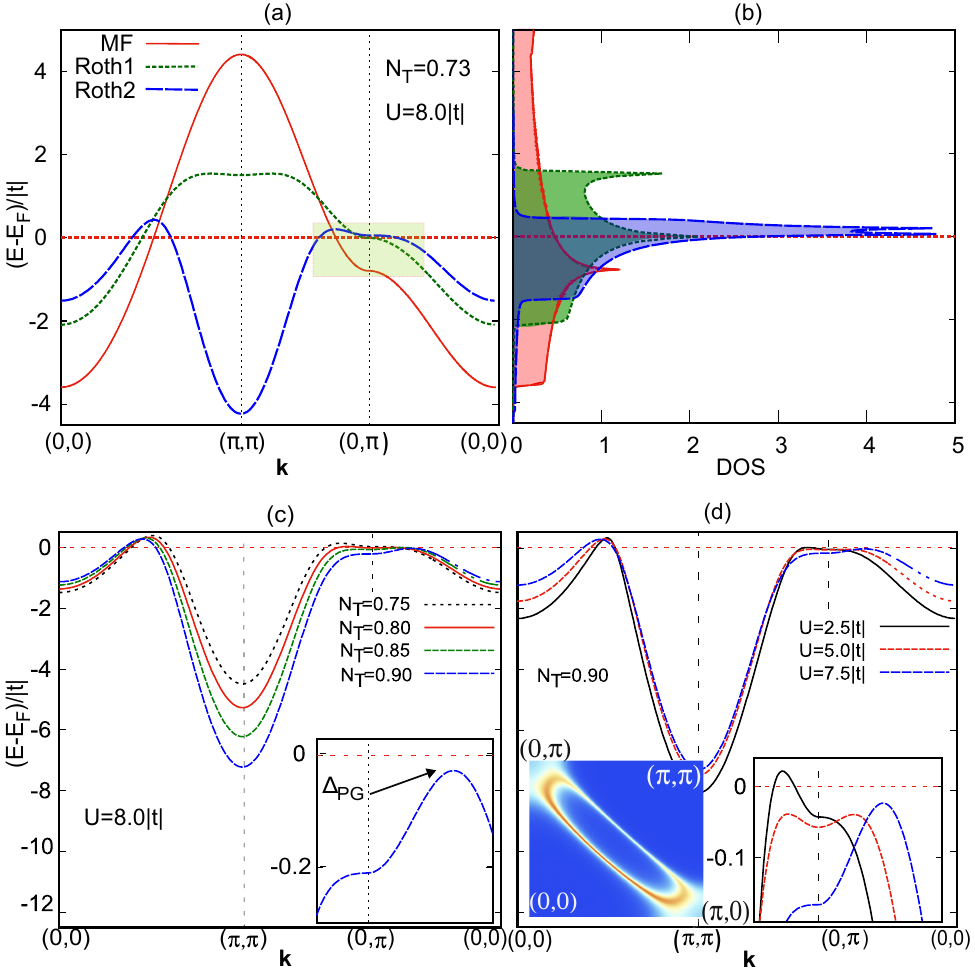}
\end{center}
\caption{ (a) The normal state quasiparticle band $E_{1\Vec{k}}$  for different levels of approximation, mean field (MF) \cite{jesus350}, Roth1
and Roth2
methods. In (b), the density of states related to the quasiparticle bands shown in (a).  The quasiparticle bands for the Roth2 method and different occupations $N_T$ in (c) and for different values of $U$ in (d). The insets in the lower right corners show in detail the opening of a pseudogap $\Delta_{PG}$ near the point $(0,\pi)$.  The inset in the lower left corner in (d), displays the spectral function $A(\vec{k},\omega=0)$. The intensity of the spectral function is minimum in blue and maximum in red. The horizontal dotted red line indicates the position of the Fermi energy $E_F$. }
\label{figbandados}
\end{figure}
Following the BCS formalism \cite{BCS}, it is considered $V_{\vec{k},\vec{k^{'}}}=-V$, therefore, $\Delta_{\vec{k}}=\Delta_{\vec{k^{'}}}=\Delta$, and then performing the summation over the Matsubara frequencies:
\begin{align}
\Delta=\frac{\Delta}{2}\frac{V}{N}\sum_{\vec{k}}\varphi(\vec{k})^2\psi(\vec{k})\left[\tanh\left(\frac{\beta \xi^{+}_{\vec{k}}}{2}\right)\alpha_{1,\vec{k}}-\tanh\left(\frac{\beta \xi^{-}_{\vec{k}}}{2}\right)\alpha_{2,\vec{k}}\right]
\label{gap}
\end{align}
with
\[ \varphi(\vec{k}) =
     \left\{
         \begin{array}{ll}
               1 & \rightarrow  \hbox{symmetry-s} \\
               \cos(k_x)-\cos(k_y) & \rightarrow \hbox {symmetry}-d_{x^2-y^2}
         \end{array}
     \right.
\]
and
\[ \psi(\vec{k}) =
     \left\{
         \begin{array}{ll}
               1 & \rightarrow  |\xi_{\vec{k}}^{\pm}|<\omega_D \\
               0 & \rightarrow \mbox{otherwise}
         \end{array}
     \right.
\]
where $\omega_D$ is the Debye frequency.
In terms of the correlated quasiparticle bands of the normal state given in Eq. (\ref{eqE1k}), the quasiparticle bands for the superconducting state are:
\begin{align}
\xi^{\pm}_{\vec{k}}=\pm\sqrt{\frac{A^{2}+B^{2}\pm\sqrt{(A^{2}+B^{2})^{2}-4A^{2}B^{2}}}{2}}
\end{align}
with $A=-E_{1,\vec{k}}$ and $B=-E_{2,\vec{k}}$. The quantities $\alpha_{1,k}$ and $\alpha_{2,k}$ are:
\begin{align}
\alpha_{1,\vec{k}}=\frac{(\xi^{+}_{\vec{k}})^{2}+(Z_{1,\vec{k}}B+Z_{2\vec{k}}A)^{2}}{2\xi^{+}_{\vec{k}}[(\xi^{+}_{\vec{k}})^{2}
-(\xi^{-}_{\vec{k}})^{2}]}
\end{align}
and
\begin{align}
\alpha_{2,\vec{k}}=\frac{(\xi^{-}_{\vec{k}})^{2}+(Z_{1,\vec{k}}B+Z_{2\vec{k}}A)^{2}}{2\xi^{-}_{\vec{k}}[(\xi^{+}_{\vec{k}})^{2}
-(\xi^{-}_{\vec{k}})^{2}]}.
\end{align}

\begin{figure*}[!h]
\begin{center}
\leavevmode
\includegraphics[angle=0,width=14cm]{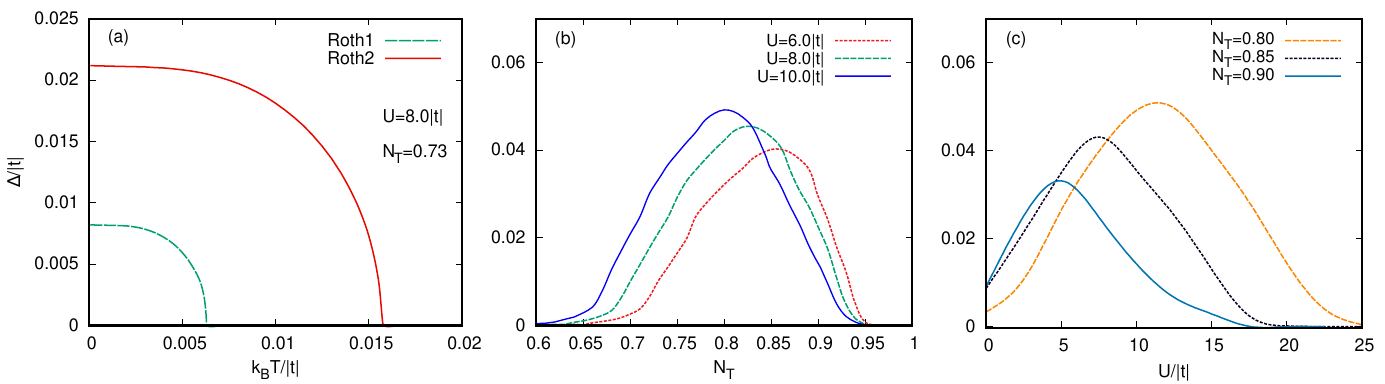}
\end{center}
\caption{ (a) The superconducting gap $\Delta$ as a function of the temperature for both, Roth1 and Roth2 methods. In (b), the superconducting gap versus the total occupation $N_T$ for different values of Coulomb interaction $U$ and  $T=0.0$ K. The behavior of $\Delta$ as a function of $U/|t|$, for different values of $N_T$, is shown in panel (c) for $T=0.0$ K. The results in (a) and (b) have been obtained with the Roth2 method for the attractive interaction $V=-0.35|t|$.  }
\label{fig_gap_d}
\end{figure*}

If we assumes that the superconducting transition is second order, close to $T_{c}$, the superconducting gap $\Delta$ tends to zero, so we can expand the gap equation around
$\Delta =0$, obtaining
\begin{align}
\frac{1}{V}=\frac{1}{2N}\sum_{\vec{k}}\varphi(\vec{k})^2\psi(\vec{k})\left[\alpha_{1,\vec{k}}\tanh\left(\frac{\beta\xi_{\vec{k}}^{+}}{2}\right)-\alpha_{2,\vec{k}}\tanh\left(\frac{\beta\xi_{\vec{k}}^{-}}{2}\right)\right],
\label{A14}
\end{align}
where $\Delta=0$ in $\alpha_{1,\vec{k}}$, $\alpha_{2,\vec{k}}$ and $\xi_{\vec{k}}^{\pm}$.

\section{Numerical results}
\label{numresults}
All the numerical results presented in this section have been obtained considering $t=-0.2$ eV, $t'=0.15|t|$ and $\omega_D=0.30|t|$.

For comparison purposes, in Fig. \ref{figbandados}(a) the normal state quasiparticle band $E_{1\Vec{k}}$ (see Eq. (\ref{eqE1k})), for the mean field  (MF) \cite{jesus350} and Roth method at different levels of approximation,
i.e.,
Roth1 \cite{beenen,Roth} and Roth2 \cite{calegari2011}, are shown. The rectangular green area highlights the region of the antinodal point $(0,\pi)$. It is important to
note that in the Roth2 method, due to electron-electron correlations, the quasiparticle band presents a larger narrow (almost flat) region around the antinodal point, when compared with the quasiparticle bands for the other approximations. As a consequence, the DOS presents a high density of states near the Fermi energy $E_F$, as can be seen in Fig. \ref{figbandados}(b) (in dark blue). A high density of states at $E_F$ provides a large number of available electrons to form pairs responsible for superconductivity, and consequently, contributes to a higher $T_c$  \cite{Kauppila,Markiewicz}.
The evolution of the quasiparticle band $E_{1\vec{K}}$ in terms of the total occupation $N_T(=n_\sigma+n_{-\sigma})$ for the Roth2 method  is shown in Fig. \ref{figbandados}(c). For $N_T$ greater than 0.85, a pseudogap emerges in the region of the antinodal point $(0,\pi)$, as shown in the inset. In the present case, the pseudogap arises due to the antiferromagnetic correlations which shift the band downward in the nodal region (point $(\pi,\pi)$), as can be seen in Figs. \ref{figbandados}(c) and \ref{figbandados}(d). Similar results have been reported in references \cite{Korshunov,Kuzmin,avella,DANTUNG}. The effect of the Coulomb interaction $U$ on the quasiparticle band $E_{1\Vec{k}}$ is shown in Fig. \ref{figbandados}(d). The inset in the lower right corner exhibits the opening of a pseudogap above a given value of $U$. The inset in the lower left corner displays the spectral function $A(\vec{k},\omega=0)$. The suppression of the spectral weight near the antinodal points indicates the presence of pseudogaps at the Fermi surface \cite{valerii,avella}.

\subsection{$d_{x_2 - y_2}$-wave pairing}

The behavior of the superconducting gap $\Delta$ as a function of the temperature for $d_{x_2 - y_2}$-wave pairing, is shown in Fig. \ref{fig_gap_d}(a). The data for the Roth1
method produces a lower gap when compared with the
Roth2 results.
This difference is mainly related to the intensity of the 
density of states at the Fermi energy DOS$_{E_F}$, which for the Roth1 result is significantly
lower than in Roth2
(see Fig.  \ref{figbandados}(b)). As discussed before, a  high DOS$_{E_F}$ should favor the superconducting state. Fig. \ref{fig_gap_d}(b) shows the gap  $\Delta$ as a function of the total occupation $N_T$ and different values of Coulomb interaction, at $T=0.0$ K. The increase in the Coulomb interaction $U$ moves the superconducting region in the direction of smaller occupancy. On the other hand, if we analyze the behavior of $\Delta$ versus $U$ for different values of $N_T$, as shown in Fig.  \ref{fig_gap_d}(c), we find that the superconducting region moves to smaller values of $U$, when $N_T$ increases. 
The combination of the results from Figs. \ref{fig_gap_d}(b) and \ref{fig_gap_d}(c), indicate that an intermediate amount of correlations is necessary for superconductivity to occur, i.e., if $U$ is very large, then the $N_T$ should decrease to reduce the intensity of the correlations. Otherwise, superconductivity is suppressed.
\begin{figure}[!h]
\begin{center}
\leavevmode
\includegraphics[angle=0,width=8.5cm]{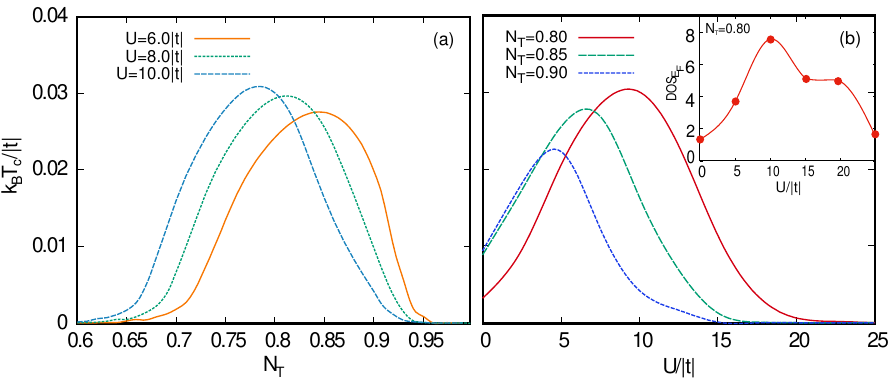}
\end{center}
\caption{In (a),  $k_BT_c$ as a function of the total occupation $N_T$ for different values of Coulomb interaction.
In (b),  $k_BT_c$ as a function of the Coulomb interaction for distinct values of $N_T$. The inset in (b), shows the behavior of the density of states at the Fermi energy, as a function of $U/|t|$, for $N_T=0.80$. These results were obtained considering the Roth2 method \cite{calegari2011,diovana}.}
\label{figTcd}
\end{figure}

The superconducting critical temperature $k_BT_c$ as a function of $N_T$ is shown in Fig. \ref{figTcd}(a) for different values of $U$, while Fig. \ref{figTcd}(b) shows $k_BT_c$ versus $U$ for different values of $N_T$. The behavior of $k_BT_c$, in both cases, is very similar to the results for the gap shown in the Figs. \ref{fig_gap_d}(b) and \ref{fig_gap_d}(c). It is interesting to note that the behavior of $k_BT_c$ as a function of the Coulomb interaction $U$ is related to the density of states at Fermi energy. As can be seen in the inset in Fig. \ref{figTcd}(b),
the DOS$_{E_F}$ increases with $U$ until reaching a maximum value at approximately the same value of $U$ in which $k_BT_c$ is also maximum. If $U$ continues to increase, both DOS$_{E_F}$ and $k_BT_c$ decrease with $U$. The result shown in Fig. \ref{figTcd}(a) is in qualitative agreement with those reported in reference \cite{Domanski}, for a boson-fermion model.
\begin{figure}
\begin{center}
\leavevmode
\includegraphics[angle=0,width=8.5cm]{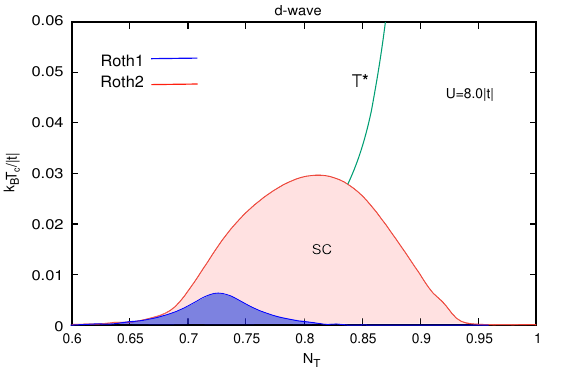}
\end{center}
\caption{The red line shows $k_BT_c$ as a function of the occupation $N_T$ for the Roth2 method while the green line indicates the temperature ($T^*$), below which, the pseudogap occurs. The dark blue line shows $k_BT_c$ for the Roth1 method.
}
\label{figTcdPG}
\end{figure}

Fig. \ref{figTcdPG} shows  $k_BT_c$ as a function of the total occupation  $N_T$ with $U=8.0|t|$: in red, the superconducting region for the Roth2 method; in dark blue, the same results for the Roth1 method. Notice that the Roth2 scheme presents superconductivity in a larger range of $N_T$
and with
a larger maximum  $k_BT_c$.
The main reason for this
is that in the Roth2 method the antiferromagnetic correlations give rise to a nearly flat band region at the antinodal points resulting in a large DOS$_{E_F}$. As a consequence, the $k_BT_c$ is enhanced in the Roth2 method.
The temperature that marks the opening of the pseudogap (T$^*$), is indicated by the green line.

\begin{figure}[!t]
\begin{center}
\leavevmode
\includegraphics[angle=0,width=6.5cm]{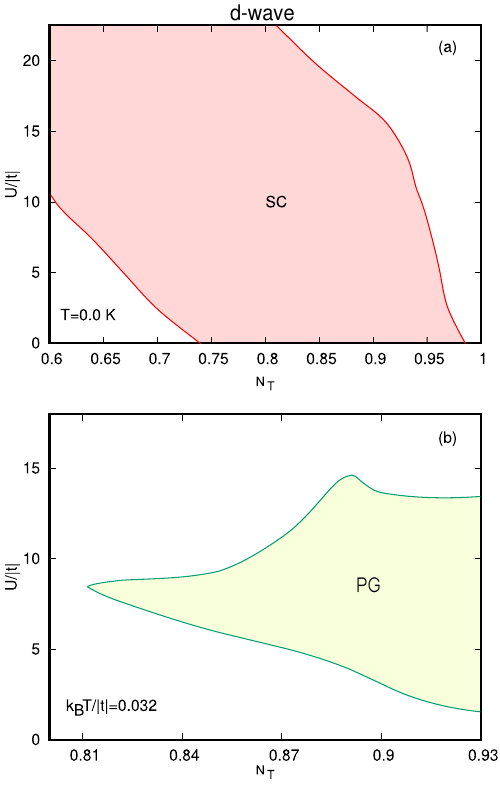}
\end{center}
\caption{In (a), the ground state diagram phase of $U$ versus $N_T$. In (b), the pseudogap region as a function of $U$ and $N_T$. These results have been obtained with the Roth2 method.
}
\label{figdf_U_NT_d}
\end{figure}

In strongly correlated electron systems described by the Hubbard model \cite{ref15}, there are two routes to increasing the electronic correlations \cite{imada}. One of them consists in
increasing the Coulomb interaction $U$, while the other consists in increasing the total occupation $N_T$. This means that we can control the level of correlations in the system by changing the parameters $U$ and $N_T$.
In order to analyze
the effect of correlations on both superconductivity and pseudogap, in Fig. \ref{figdf_U_NT_d}(a) we show the ground state diagram phase for $U$ versus $N_T$, while Fig. \ref{figdf_U_NT_d}(b), exhibits the pseudogap region for $k_BT/|t|=0.032$.  In Fig.
\ref{figdf_U_NT_d}(a), it is interesting to note that
a minimum value of total occupation $N_T$ is necessary for superconductivity to emerge. On the other hand, superconductivity is present even for $U=0.0$. Indeed, the $U=0.0$ case obtains the result from the BCS model \cite{BCS}. For finite values of $U$, the superconductivity occurs in a range of $U$ and $N_T$ associated to an intermediate level of electronic correlations.
Note that as $U$ increases, $N_T$ must  decrease in order to maintain this intermediate level of correlation.
For large values of both $U$ and $N_T$,  superconductivity is suppressed due to the strong correlations. The emergence  of the pseudogap also depends on $U$ and $N_T$, as shown in Fig. \ref{figdf_U_NT_d}(b). The pseudogap emerges for large occupations. Moreover, the range of $U$ in which the pseudogap occurs increases with $N_T$, while in the superconducting case, the upper limit of $U$ decreases with $N_T$. The difference between the results in Figs.\ref{figdf_U_NT_d}(a) and \ref{figdf_U_NT_d}(b) occurs due to the fact that  superconductivity depends on how the Coulomb interaction affects the density of states at the Fermi energy, while the pseudogap depends on how $U$ distorts the quasiparticle band $E_{1\vec{k}}$ near the antinodal points. As can be seen in the inset at the lower right corner of Fig. \ref{figbandados}, for small values of $U$, the quasiparticle band crosses the Fermi energy near the point $(0,\pi)$. The same occurs above a given value of $U$. Therefore, the pseudogap only occurs in an intermediate range of $U$, which also depends on $N_T$.

\subsection{$s$-wave pairing}

\begin{figure}
\begin{center}
\leavevmode
\includegraphics[angle=0,width=8.5cm]{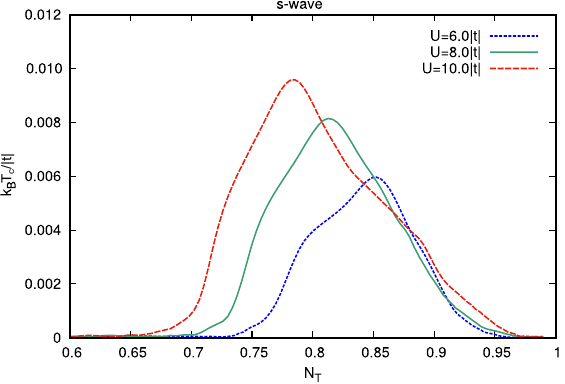}
\end{center}
\caption{  $k_BT_c$ as a function of occupation $N_T$ with different values of $U$, for the Roth2 method. }
\label{figTcs}
\end{figure}

Superconductivity with $s$-wave pairing was also considered in the present work. The $k_BT_c$ dependence on $N_T$ for different values of $U$ is shown in Fig. \ref{figTcs}. For the considered $U$ values,
the increase
 of the coulomb interaction $U$ favors  superconductivity with $s$-wave pairing. Furthermore, the range of $N_T$ in which superconductivity occurs, is enlarged by $U$. The maximum $k_BT_c$ is also enhanced by the Coulomb interaction $U$.

\begin{figure}[!h]
\begin{center}
\leavevmode
\includegraphics[angle=0,width=7.5cm]{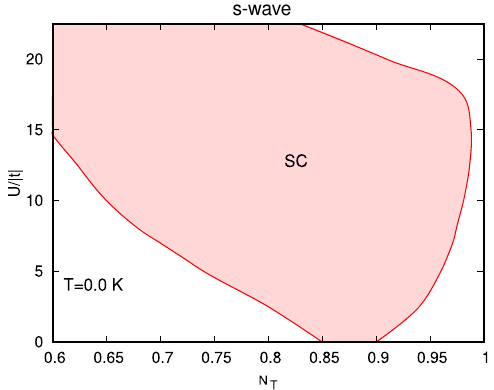}
\end{center}
\caption{ The ground state diagram phase of $U$ versus $N_T$, for the Roth2 method. }
\label{figdf_U_NT_s}
\end{figure}

\begin{figure}
\begin{center}
\leavevmode
\includegraphics[angle=0,width=7.5cm]{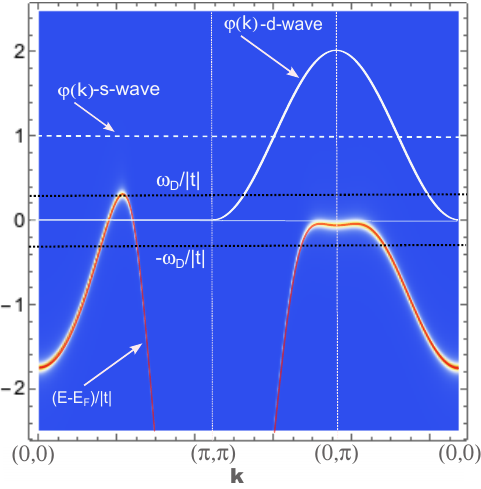}
\end{center}
\caption{ In red, the quasiparticle band $E_{1\vec{k}}$ for the Roth2 method in the normal state. The model parameters are $U=5.0|t|$ and $N_T=0.90$, while $k_BT=0.0$. The solid line in white shows the function $\varphi(\vec{k})$ for the $d$-wave case while the dashed line in white shows  the function $\varphi(\vec{k})$ for the $s$-wave case. The horizontal dotted black lines show the cutoff frequency $\omega_D/|t|=0.3$.
}
\label{figsymmetry}
\end{figure}

The ground state diagram phase of $U$ versus $N_T$, is shown in Fig. \ref{figdf_U_NT_s}. For $U=0.0$,  superconductivity occurs only in a small range of $N_T$. However, for finite $U$, the range of $N_T$ increases with $U$ untill $U/|t|\approx 15$, otherwise, for $U/|t|\gtrsim 15$, the increase of $U$ starts to inhibit the superconductivity in the underdoped regime.
Comparing the ground state diagram phase of $U$ versus $N_T$ for both symmetries, $d_{x_2 - y_2}$-wave (Fig. (\ref{figdf_U_NT_d}(a)) and $s$-wave,
the superconducting region reaches lower values of both $U$ and $N_T$, in the $d_{x_2 - y_2}$-wave case. Comparing the maximum $k_BT_c$ for the s-wave symmetry (see Fig. \ref{figTcs}) with that of the $d$-wave symmetry, we verify that the $d$-wave pairing (see Fig. \ref{figTcd}) presents a larger maximum $k_BT_c$.
Furthermore, the maximum $k_BT_c$ for the $s$-wave symmetry (see Fig. \ref{figTcs}), is significantly smaller than the maximum $k_BT_c$ for the $d$-wave symmetry (see Fig. \ref{figTcd}(a)).
This difference can be better understood by analyzing Fig. \ref{figsymmetry} which exhibits the quasiparticle band $E_{1\vec{k}}$ and the function $\varphi(\vec{k})$ (defined after Eq. (\ref{gap})), at the high symmetry directions on the first Brillouin zone.
The width of the $E_{1\vec{k}}$ curve indicates the intensity of the spectral weight for each value of the $\vec{k}$-vector.
Considering the $d$-wave case, the function $\varphi(\vec{k})$ is null in the direction $(0,0)$-$(\pi,\pi)$, but it is maximum at the antinodal point $(0,\pi)$, where the quasiparticle band $E_{1\vec{k}}$ is nearly flat and has a high spectral weight. Moreover, due to the cutoff frequency $\omega_D$, the main contribution for the sum over $\vec{k}$ in Eqs. (\ref{gap}) and (\ref{A14}), comes from the region near the antinodal points,
where $\varphi(\vec{k})_{d-wave}\approx 2\varphi(\vec{k})_{s-wave}$. Considering that the Eqs.  (\ref{gap}) and (\ref{A14}) depend on $\varphi^{2}(\vec{k})$, thus  $\varphi^{2}(\vec{k})_{d-wave}\approx 4\varphi^{2}(\vec{k})_{s-wave}$. Therefore, the gap $\Delta$ for the $d$-wave case is significantly larger than the gap for the $s$-wave case. This is the main reason
that
the $d$-wave superconductivity is more robust
with regard to the effects of $U$ and $N_T$ and presents a larger maximum $k_BT_c$, in comparison to $s$-wave symmetry.

\begin{figure}[!ht]
\begin{center}
\leavevmode
\includegraphics[angle=0,width=6cm]{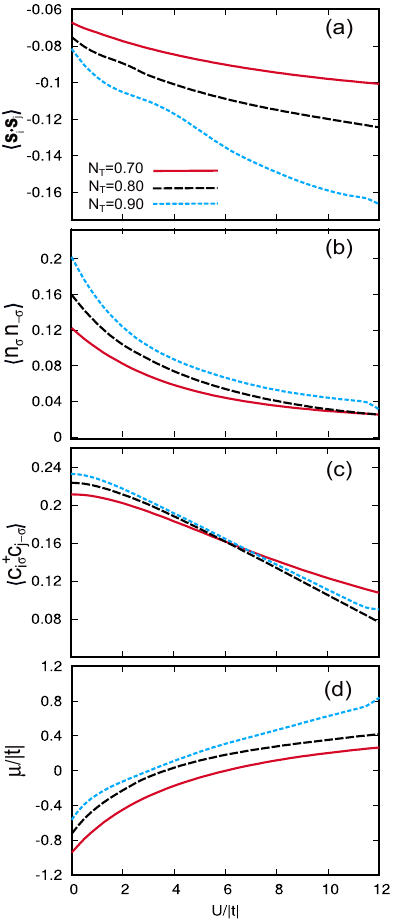}
\end{center}
\caption{ (a) The spin-spin correlation function $\langle \vec{S}_i\cdot \vec{S}_j\rangle$ versus the Coulomb interaction $U$ for different occupations $N_T$. In (b), the double occupation, in (c) the hopping correlation function and in (d), the chemical potential as a function of $U$. These results have been obtained for the normal state considering the Roth2 method with $k_BT=0.032|t|$.  }
\label{figSiSj}
\end{figure}

The behavior of some important correlation functions as a function of the Coulomb interaction is shown in Fig. \ref{figSiSj}. In Fig. \ref{figSiSj}(a), the amplitude of the spin-spin correlation function $\langle \vec{S}_i\cdot \vec{S}_j\rangle$ is shown for different values of $N_T$. Because $\langle \vec{S}_i\cdot \vec{S}_j\rangle$ is negative, it is related to antiferromagnetic correlations \cite{beenen}, therefore, it can be concluded that the antiferromagnetic correlations are enhanced by both, $U$ and $N_T$.
As discussed before, the antiferromagnetic correlations
related to the spin-spin correlation function $\langle \vec{S}_i\cdot \vec{S}_j\rangle$ are the source of a pseudogap in the present scenario \cite{Korshunov,Kuzmin,avella,DANTUNG}.
Fig. \ref{figSiSj}(b) displays the double occupation $\langle n_{\sigma}n_{-\sigma}\rangle$ as a function of $U$. Notice that the double occupation decreases with $U$, but for lower occupancy values, $\langle n_{\sigma}n_{-\sigma}\rangle$ is less affected by $U$. This result is in agreement with recent results for the Hubbard model \cite{Fedor}.
The hopping correlation function $\langle c_{i\sigma}^{\dagger}c_{j\sigma}\rangle$ is shown in Fig. \ref{figSiSj}(c). It is clear that $\langle c_{i\sigma}^{\dagger}c_{j\sigma}\rangle$ decreases with increasing $U$ and the effect of $U$ on  $\langle c_{i\sigma}^{\dagger}c_{j\sigma}\rangle$  is more intense near to half filling. A similar behavior has been reported for the Hubbard model in three dimensions, using a similar technique \cite{Nolting}. Finally, in Fig. \ref{figSiSj} (d), we present the chemical potential as a function of $U$. Initially, the chemical potential increases with $U$ in order to control the occupation per site of the lattice. For large values of $U$ (not shown), all sites are singly occupied, therefore the chemical potential becomes independent of $U$. The results for the chemical potential are in qualitative agreement with that reported in reference \cite{Fedor} for the Hubbard model.

\section{Conclusions}
\label{conclusions}

The interesting interplay between pseudogap and superconductivity was investigated in this work.
In order to address this issue,  Green's functions for the one-band Hubbard model
were treated within the n-pole approximation proposed by Roth \cite{Roth}. However, in the present work, it has been considered an improved version of the n-pole approximation, called here, the Roth2 method \cite{calegari2011}.
This method allows an
investigation of the effects of antiferromagnetic correlations on both the pseudogap and superconductivity. Indeed, the antiferromagnetic correlations associated with the spin-spin correlation function  $\langle \vec{S}_i\cdot \vec{S}_j\rangle$, move the quasiparticle band to lower energies in the region of the nodal point $(\pi,\pi)$. As a consequence, a nearly flat region appears around the antinodal point $(0,\pi)$.  When the flat region is completely below the Fermi energy in the normal state, a pseudogap emerges at the antinodal points of the Fermi surface. In the superconducting case, if the nearly flat region is close to the Fermi energy $E_F$,  it causes a large density of states at $E_F$ favoring  superconductivity and consequently, enhancing the superconducting critical temperature. Results highlighting this relation between the flat bands and superconductivity are reported in references \cite{aoki,edwin,edwin1,Kauppila,Markiewicz,Melnikov}. On the other hand, the emergence of a pseudogap due to antiferromagnetic correlations, is discussed in references  \cite{Harrison,Morinari,valerii,Korshunov,Kuzmin,avella}.

Considering that the correlations in the system become strong when the Coulomb interaction and/or the total occupation increase, we verified that  superconductivity occurs at an intermediate correlation level, in which the quasiparticle bands present a nearly flat band region at the antinodal points.

In summary, the main achievement of the present work is to present a scenario in which the antiferromagnetic correlations play an important role in a strongly correlated electron system.
The present results show that
the antiferromagnetic correlations may affect the quasiparticle band structure, giving rise to nearly flat band regions at the antinodal points. As a consequence of this feature, a pseudogap emerges in the normal state
and the superconducting critical temperature is increased in the superconducting state.

\section*{Acknowledgements}
This work was partially supported by Coordena\c{c}\~ao de Aper\-fei\c{c}oa\-mento de Pessoal de N\'{\i}vel Superior (CAPES); Conselho Nacional de Desenvolvimento Cient\'ifico
e Tecno\-l\'o\-gico  (CNPq), 309598/2020-6 (R.L.S.F.);
Funda\c{c}\~ao de Amparo \`a Pesquisa do Estado do Rio Grande do Sul (FAPERGS), Grants Nos. 19/2551- 0000690-0
and 19/2551-0001948-3 (R.L.S.F.); The work is also part of the
project Instituto Nacional de Ci\^encia e Tecnologia - F\'isica Nuclear e Aplica\c{c}\~oes
(INCT - FNA), Grant No. 464898/2014-5.

\end{document}